\documentclass[onecolumn,amsmath,amssymb,superscriptaddress,nofootinbib,11pt,a4paper]{revtex4}
\usepackage{titlesec}
\titleformat{\section}
  {\normalfont\Large\bfseries}  
  {\thesection}                 
  {1em}                         
  {}                            
\titleformat{\subsection}
  {\normalfont\large\bfseries}
  {\thesubsection}
  {1em}
  {}
\pdfoutput=1
\usepackage[T1]{fontenc}
\usepackage{CJK}
\usepackage{xcolor}
\usepackage{amsfonts}
\usepackage{epstopdf}
\usepackage{wrapfig} 
\usepackage{graphicx}  
\usepackage{dcolumn}   
\usepackage{bm}
\usepackage{float}
\usepackage[utf8]{inputenc}
\usepackage{booktabs} 
\usepackage{multirow}
\usepackage{subfigure}

\begin{document}

\title{Draining Without Emptying a Stringy Black Hole: Repetitive Electric Penrose Process}

\author{Huai-Fan Li}\email{huaifan999@163.com (Corresponding author)}
\affiliation{College of General Education, Shanxi College of Technology, Shuozhou, 036000, China}
\author{Xiong-Ying Guo}
\email{g1x1y1@126.com}
\affiliation{Department of Physics, Shanxi Datong University, Datong, 037009, China}
\author{Ke Wang}
\email{kkwwang2025@163.com}
\affiliation{School of Material Science and Engineering, Chongqing Jiaotong University, Chongqing 400074, China}
\author{Hong Lei}
\email{405114039@qq.com}
\affiliation{Chongqing Bashu Science City  Secondary School,Chongqing 401331, China}

\begin{abstract}
This paper studies the repetitive electric Penrose process in the Gibbons-Maeda-Garfinkle-Horowitz-Strominger (GMGHS) black hole under the optimal condition for maximum energy extraction. Unlike the Reissner-Nordström (RN) black hole, the effective potential curve of the incident particle in the GMGHS black hole remains unchanged during the iterations. The iteration termination conditions are determined by two aspects: the escape condition requires that particle 0 can overcome the effective potential peak and escape to infinity, and this condition is independent of the black hole charge; the existence condition for negative-energy particle 1 requires the black hole charge to be sufficiently large, indicating that the repetitive process cannot completely consume the charge of the GMGHS black hole to zero. Numerical simulations show that the return on investment of energy extraction can exceed $100\%$, but the energy utilization efficiency does not exceed $50\%$, and most of the extractable energy is converted into an increase in the irreducible mass of the black hole. In addition, a smaller decay radius and a larger absolute value of the charge-to-mass ratio of particle 1 can improve the return on investment and energy utilization efficiency of energy extraction; the larger the specific energy of the incident particle, the smaller the return on investment of energy extraction, while the energy utilization efficiency remains unchanged; for an initially non-extremal black hole, as the initial charge increases, the return on investment of energy extraction increases monotonically, while the energy utilization efficiency first increases and then decreases. After the iterations end, the black hole still retains extractable energy, and neither the charge nor the mass vanishes.
\end{abstract}

\maketitle
 \newpage
\section{Introduction}
Black hole energy extraction is one of the most important frontier topics in general relativity and black hole thermodynamics. As a classical energy extraction mechanism\cite{16}, the Penrose process was first proposed by Penrose in 1969. Its basic idea is to split an incident particle in the ergosphere of a black hole, so that one part of the particle falls into the black hole (carrying negative energy), while the other part escapes to infinity with positive energy larger than the incident energy. In 1973, Misner, Thorne, and Wheeler proposed a repetitive Penrose process\cite{17}, arguing that through repeated iterations, all extractable energy could be extracted. Until 2025, Ruffini et al.\cite{7,8} found that the repetitive Penrose process cannot extract all the extractable energy (for related early studies, see \cite{18}). This is because the iterative process is nonlinear; at each iteration, the mass and angular momentum of the black hole need to be updated, and the irreducible mass also increases nonlinearly. Based on the repetitive Penrose process discovered by Ruffini et al., it has now been generalized to many spacetime backgrounds\cite{9,15,19,20,21,22,23,24}.

In the above literature, the Penrose process used is the mechanical Penrose process, which extracts the rotational energy of the black hole. In fact, four years after the Penrose process was proposed, i.e., in 1973, Denardo and Ruffini discovered the electric Penrose process\cite{25}. Because negative-energy orbits exist in the generalized ergosphere, the energy extracted at this time is the electric energy of the black hole. Similarly, the electric Penrose process can still be iterated, for example, in black hole bombs\cite{26,27,5}. However, because these repetitive electric Penrose processes did not update the mass and charge of the black hole, the waste in the extraction process was not considered. Recently, Hu et al.\cite{6}, following the work of \cite{7,8}, corrected the repetitive electric Penrose process by performing nonlinear iterations. They found that the charge and extractable energy of an RN black hole cannot be reduced to zero through a finite number of iterations. Alipour et al.\cite{28}, building on the work of Hu et al., generalized the electric Penrose process to Kerr-Newman black holes and investigated the repetitive Penrose process under the co-evolution of charge and spin.

However, charged black holes are not only RN black holes. String theory is a theoretical framework that regards the fundamental constituents of the universe not as point-like particles but as different vibrational modes of extremely small one-dimensional ``strings,'' and attempts to unify quantum mechanics and general relativity\cite{29,30}. String theory, as a leading candidate for a self-consistent quantum gravity framework, possesses a low-energy effective description in four-dimensional spacetime known as the Einstein-Maxwell-Dilaton-Axion (EMDA) theory\cite{12}. The intricate coupling of the dilaton and axion fields with the electromagnetic sector endows this theory with rich intrinsic symmetries and distinctive black hole quantum properties\cite{31,32}. Extensive literature has explored various facets of the EMDA framework\cite{33,34,35,36,37}. Consequently, probing the astrophysical signatures predicted by EMDA theory offers a compelling avenue to test gravity theories beyond general relativity. The charged black hole solution within the Einstein-Maxwell-dilaton model was initially formulated by Gibbons and Maeda\cite{1}. This solution was later reformulated by Garfinkle, Horowitz, and Strominger within the context of string theory\cite{2}, yielding the static, spherically symmetric GMGHS solution. On this basis, Sen generalized it to a rotating version, namely the Kerr-Sen solution\cite{12}. When the black hole is uncharged and non-rotating, the metric reduces to the Schwarzschild solution. Many properties of the GMGHS black hole have been extensively studied\cite{3,13,14,38,39}. Due to the differences in the metric, the particle dynamics, effective potential structure, and evolutionary features of energy extraction in the GMGHS black hole may exhibit new physical phenomena different from those of the standard RN black hole. At present, a systematic study of the repetitive electric Penrose process in the GMGHS black hole is still lacking. This paper aims to fill this gap and systematically study the repetitive electric Penrose process in the GMGHS black hole, and explore its differences from the RN black hole.

The structure of this paper is as follows: In Section 2, we introduce the electric Penrose process in the GMGHS black hole; in Section 3, we derive the optimal condition for maximum energy extraction; in Section 4, we give in detail the iteration termination conditions and determine the allowed region for energy extraction; in Section 5, we present numerical simulation results and discuss the return on investment, energy utilization efficiency, and parameter dependence of energy extraction; finally, in Section 6, we give our conclusions and outlook.

\section{Electric Penrose Process in the GMGHS Black Hole}
In Boyer-Lindquist coordinates $(t, r, \theta, \phi)$, the GMGHS metric is\cite{1,2,3}
\begin{equation}
ds^{2} = -\left(1 - \frac{2M}{r}\right)dt^{2} + \left(1 - \frac{2M}{r}\right)^{-1}dr^{2}
+ r\left(r - \frac{Q^{2}}{M}\right)d\theta^{2} + r\left(r - \frac{Q^{2}}{M}\right)\sin^{2}\theta\, d\phi^{2}.
\end{equation}
Here, $M$ is the black hole mass and $Q$ is the black hole charge. The event horizon is located at $r=2M$. The charge range is $0\leq Q^{2}\leq 2M^2$, and the extremal black hole is at $Q= \sqrt{2}M$. The electromagnetic potential of the black hole is $A_t=-\frac{Q}{r}$. According to the charge range, we obtain $0\leq Q^{2}/M\leq 2M$. Outside the horizon, $r>2M$, so $r>Q^2/M$ is strictly satisfied; on the horizon and in the extremal black hole case, $r=Q^2/M$.

We consider the electric Penrose process in this black hole, in which particle 0 splits into particle 1 and particle 2. The conservation laws of four-momentum and charge are
\begin{equation}
\hat{E}_0 = \tilde{\mu}_1 \hat{E}_1 + \tilde{\mu}_2 \hat{E}_2,\quad
\hat{q}_0 = \tilde{\mu}_1 \hat{q}_1 + \tilde{\mu}_2 \hat{q}_2,\quad
\hat{P}_0 = \tilde{\mu}_1 \hat{P}_1 + \tilde{\mu}_2 \hat{P}_2,    \label{2}
\end{equation}
where 
\begin{equation}
\hat{E}_i = E_i / \mu_i,\quad\hat{q}_i = q_i / \mu_i,\quad\hat{P}_i = P_i / \mu_i ,\quad
\tilde{\mu}_i = \mu_i / \mu_0.
\end{equation}
Here, $E_i,\mu_i,q_i,P_i$ represent the energy, mass, charge, and radial momentum of particle $i(i=0,1,2)$, respectively. Recall from Refs.\cite{4,5,6} that particle 0 starts from infinity and is made static or has an initial radial velocity, so that particle 1 or particle 2 has radial velocity, thereby satisfying the optimal condition for maximum energy extraction. The particles undergo purely radial motion, with four-velocity $u^\mu = (\dot{t}, \dot{r}, 0, 0)$, where the dot denotes the derivative with respect to proper time $\tau$. The conserved energy is
\begin{equation}
E = - (p_t + q A_t)=-g_{tt} p^t-q A_t=-\mu g_{tt} \dot{t} - q A_t = \mu \left(1 - \frac{2M}{r}\right)\dot{t} + \frac{q Q}{r}.
\end{equation}
From the normalization of the four-velocity 
\begin{equation}
u^\mu u_\mu = -\left(1 - \frac{2M}{r}\right) \dot{t}^2 + \left(1 - \frac{2M}{r}\right)^{-1} \dot{r}^2 = -1,
\end{equation}
we obtain
\begin{equation}
\dot{r}^2 = \left(\hat{E} - \frac{\hat{q} Q}{r}\right)^2 - \left(1 - \frac{2M}{r}\right).
\end{equation}
The dimensionless quantities are
\begin{equation}
\hat{Q} = Q/M, \quad\hat{P} = \dot{r} ,\quad\hat{r}=r/M,
\end{equation}
so
\begin{equation}
\hat{P}_i^2 = (\hat{E}_i - \hat{V}_i^-)(\hat{E}_i - \hat{V}_i^+), \quad \hat{V}_i^{\pm} = \frac{\hat{Q}\hat{q}_i}{\hat{r}} \pm \sqrt{1 - \frac{2}{\hat{r}}}.\label{8}
\end{equation}
Here, $\hat{V}_i^{\pm}$ are the effective potentials for radial motion.

\section{Optimal Condition for Maximum Energy Extraction}
References\cite{6,7,8,9} show that in the Penrose process, the optimal condition for maximum energy extraction corresponds to the case where the decay occurs when all three particles are at their own turning points and have zero kinetic energy, i.e., the radial momenta of the three particles must vanish at the decay location. We will prove below that this correspondence still holds in the electric Penrose process in the GMGHS black hole. Since $\hat{P}_i^2\geq0$, we have $\hat{E}_i \geq \hat{V}_i^+$ or $\hat{E}_i \leq \hat{V}_i^-$. From the time component of the four-velocity, combined with the conserved energy expression, we obtain
\begin{equation}
\dot{t} = \frac{\hat E_i - \hat Q \hat q_i/\hat r}{1 - 2/\hat r}.
\end{equation}
Outside the horizon \(\hat r>2\), the denominator is positive, so the timelike condition \(\dot{t}> 0\) is equivalent to
\begin{equation}
\hat E_i - \frac{\hat Q \hat q_i}{\hat r} > 0.\label{10}
\end{equation}
If we take the negative branch \(\hat E_i \le \hat V_i^-\), i.e.,
\begin{equation}
\hat E_i \le \frac{\hat Q \hat q_i}{\hat r} - \sqrt{1 - \frac{2}{\hat r}},
\end{equation}
then
\begin{equation}
\hat E_i - \frac{\hat Q \hat q_i}{\hat r} \le -\sqrt{1 - \frac{2}{\hat r}} < 0,
\end{equation}
which contradicts Eq.\eqref{10}. Therefore, only the positive branch is physically allowed:
\begin{equation}
\hat E_i \ge \hat V_i^+.
\end{equation}
In order to achieve energy extraction, particle 1 satisfies $\hat E_1 < 0$ and $\hat E_1 \ge \hat V_1^+$. To maximize the extracted energy, \(\hat E_1\) should be as small as possible (as negative as possible), so its lower bound must be saturated, i.e., $\hat E_1 = \hat V_1^+$. Substituting particle 1 into the radial motion equation gives
\begin{equation}
\hat P_1^2 = (\hat V_1^+ - \hat V_1^-)(\hat V_1^+ - \hat V_1^+) = 0.
\end{equation}
Then according to the third equation of Eq.\eqref{2}, we obtain $P_0=P_2$. We assume $P_0=P_2\neq0$; from Eq.\eqref{8} we obtain
\begin{equation}
E_0 - V_0^- = E_2 - V_2^-,\quad E_0 - V_0^+ = E_2 - V_2^+.
\end{equation}
Then
\begin{equation}
E_0 - \frac{\hat{Q} q_0}{\hat{r}} = E_2 - \frac{\hat{Q} q_2}{\hat{r}}.
\end{equation}
From the first two equations of Eq.\eqref{2}, we further obtain
\begin{equation}
\hat{E}_1 = \frac{\hat{Q}\hat{q}_1}{\hat{r}}.
\end{equation}
Obviously, this violates the timelike condition\eqref{10}, indicating that we must have $P_0=P_2=0$. Therefore, under the optimal condition for maximum energy extraction, the radial momenta of the three particles must be zero. Denoting the dimensionless decay radius by $r_d$, the solution of the basic equations\eqref{2} for the electric Penrose process in the GMGHS black hole, given $\hat{E}_0,\hat{q}_1, \nu = \mu_2 / \mu_1$, is
\begin{equation}
    \begin{aligned}
&\tilde{\mu}_1 = \frac{1}{1+\nu},\quad \tilde{\mu}_2 = \frac{\nu}{1+\nu},\quad\hat{E}_1 = \frac{\hat{Q}\hat{q}_1}{\hat r_d} + \sqrt{1-\frac{2}{\hat r_d}},\\
&\hat{q}_0 = \frac{\hat r_d}{\hat Q}\left(\hat{E}_0 - \sqrt{1-\frac{2}{\hat r_d}}\right),\quad\hat{q}_2 = \frac{\hat{q}_0 - \tilde{\mu}_1 \hat{q}_1}{\tilde{\mu}_2},\quad\hat{E}_2 = \frac{\hat{Q}\hat{q}_2}{\hat r_d} + \sqrt{1-\frac{2}{\hat r_d}}.
\end{aligned}\label{18}
\end{equation}
Equation\eqref{18} differs from the solution of the basic equations for the electric Penrose process in the Reissner-Nordstrom black hole in Ref.\cite{6} (their equations 10 to 14), due to the different metric.

\section{Iteration Termination Conditions}
After each energy extraction, the remaining mass and charge of the black hole are
\begin{equation}
M_n = M_{n-1} + \mu_0 \tilde{\mu}_1 \hat{E}_{1,n-1},\quad
Q_n = Q_{n-1} + \mu_0 \tilde{\mu}_1 \hat{q}_1.  
\end{equation}
The iteration process cannot continue indefinitely; it needs to satisfy the following two conditions.

First, particle 1 must fall into the black hole, while particles 0 and 2 must escape to infinity. For the negatively charged particle 1, differentiating the effective potential of particle 1 with respect to the decay radius shows that outside the event horizon the derivative is always greater than 0, i.e., $\hat{V}_1^+(\hat{r}_d)$ is monotonically increasing. Since $\hat{E}_1 = \hat{V}_1^+(\hat{r}_d)$, and $\hat{V}_1^+$ decreases as $\hat{r}$ decreases, when particle 1 moves from $\hat{r}_d$ toward the event horizon, we have
\begin{equation}
 \hat{P}_1^2 = (\hat{E}_1 - \hat{V}_1^-)(\hat{E}_1 - \hat{V}_1^+) > 0,   
\end{equation}
i.e., particle 1 is allowed to continue moving inward and will not encounter a classical turning point, indicating that particle 1 can always cross the event horizon. For particles 0 and 2, their turning points must lie to the right of their respective effective potential peaks. The critical condition for each particle is that its turning point coincides with the effective potential peak, i.e.,
\begin{equation}
\hat{V}_i^+(\hat{r}_d) = \hat{E}_i , \quad d\hat{V}_i^+/d\hat{r}|_{\hat{r}=\hat{r}_d} = 0.   
\end{equation}
Solving gives
\begin{equation}
\hat Q \hat q_i = \frac{1}{\sqrt{1-\frac{2}{\hat r_d}}}, \quad    \hat E_i= \frac{1}{\hat r_d\sqrt{1-\frac{2}{\hat r_d}}}+ \sqrt{1-\frac{2}{\hat r_d}}.
\end{equation}
Since $\hat E_1 < 0$, energy conservation gives $\hat E_2 > \hat E_0$. Therefore, the escape condition for iteration termination is determined by particle 0:
\begin{equation}
\hat E_0 \ge \sqrt{1-\frac{2}{\hat r_d}} + \frac{1}{\hat r_d \sqrt{1-\frac{2}{\hat r_d}}}   ,\label{23}
\end{equation}
This shows that the first iteration termination condition is independent of the black hole charge $\hat Q$. This can also be understood from another aspect: after knowing $\hat E_0$ and $\hat r_d$, the effective potential of particle 0 is uniquely determined:
\begin{equation}
V_0^+(\hat r)=
\frac{\hat r_d}{\hat r}
\left(
\hat E_0 - \sqrt{1-\frac{2}{\hat r_d}}
\right)+
\sqrt{1-\frac{2}{\hat r}},  
\end{equation}
independent of the iteration number. For the RN black hole, in Fig. 2 of Ref.\cite{6}, they plotted the effective potential as a function of the decay radius for different iteration numbers; in the GMGHS black hole, this need not be plotted, because no matter how many iterations are performed, there is only one effective potential curve, which remains unchanged. When $\hat E_0 = 1$, i.e., the incident particle is released from rest at infinity, $\hat r_d =\infty$, indicating that this is forbidden for the electric Penrose process in the GMGHS black hole.

Second, at each iteration, $\hat E_1 < 0$ must be strictly satisfied, which is equivalent to requiring
\begin{equation}
\hat{Q} > \frac{\sqrt{\hat r_d(\hat r_d - 2)}}{|\hat{q}_1|}.\label{25} 
\end{equation}
Since $r_d\geq2$, we have $\hat{Q}>0$, which preliminarily indicates that the repetitive electric Penrose process in the GMGHS black hole cannot consume all the charge of the black hole. However, by increasing $|\hat{q}_1|$, one can make $\hat{Q}\approx0$. For the RN black hole\cite{6}, by increasing $|\hat{q}_1|$, one can also make $\hat{Q}\approx0$ (see their Fig. 1). However, neither black hole can be said to have its charge completely reduced to zero.

\begin{figure}[!h]
  \centering
    \includegraphics[width=0.9\linewidth]{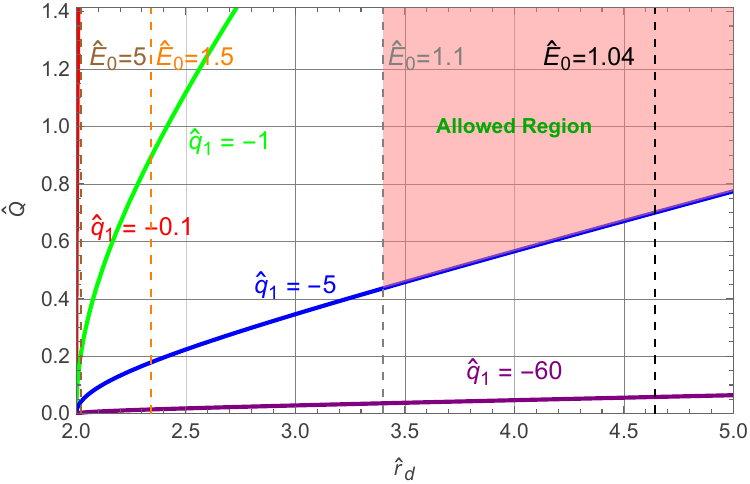} 
 \caption{Allowed region for energy extraction plotted according to the iteration termination conditions for energy extraction. The allowed region lies above the solid line and to the right of the dashed line.}
\label{fig:1}
\end{figure}

Under the influence of the iteration termination conditions\eqref{23} and\eqref{25}, given $\hat{E}_0$ and $\hat{q}_1$, the decay radius and black hole charge for energy extraction are strictly constrained. In Fig.\ref{fig:1}, we plot the allowed region for energy extraction in the $\hat{Q}-\hat{r}_d$ plane.
In addition, since there are a large number of microscopic particles with very large charge-to-mass ratios in the universe, after the electric Penrose process, the charge of the GMGHS black hole is approximately zero, but cannot disappear completely.

\section{Energy Extraction Efficiency and the Repetitive Electric Penrose Process}
The horizon area of the black hole is
\begin{equation}
S = \int_0^\pi \int_0^{2\pi} \sqrt{g_{\theta\theta} g_{\phi\phi}}  d\phi d\theta =8\pi (2M^2-Q^2).
\end{equation}
The vanishing horizon area of an extremal GMGHS black hole does not mean that the horizon disappears, but rather that it degenerates into a zero-area singular horizon; the singularity coincides with the horizon and does not constitute a naked singularity\cite{2}. For each iteration, the irreducible mass of the black hole is\cite{10}
\begin{equation}
M_{irr,n} = \sqrt{\frac{S}{16\pi}}=\sqrt{M_n^2-\frac{Q_n^2}{2}}.
\end{equation}
The extractable energy of the black hole is
\begin{equation}
E_{extractable,n} = M_n - M_{irr,n} = M_n -\sqrt{M_n^2-\frac{Q_n^2}{2}}.
\end{equation}
For an extremal GMGHS black hole, $Q= \sqrt{2}M$, the irreducible mass of the black hole is 0, and all of the black hole mass is extractable. However, we will see later that we cannot extract all of the black hole mass. The return on investment of energy extraction is\cite{6}
\begin{equation}
\xi_n =  \frac{E_{extracted,n}}{n E_0}.  
\end{equation}
The energy utilization efficiency is\cite{6}
\begin{equation}
\Xi_n = \frac{E_{extracted,n}}{E_{extractable,0} - E_{extractable,n}} .
\end{equation}

We track a complete iterative sequence of the repetitive Penrose process in Table\ref{tab:penrose}. The initial black hole is set to be extremal. Following Ref.\cite{8}, we take $\mu_0 = 0.01M_0, \nu = 0.78345$; following Ref.\cite{6}, we take $\hat{E}_0 = 1.1, \hat{q}_1 = -5$. The allowed region is located in the light red region in Fig.\ref{fig:1}, and we take the decay radius $\hat r_d=4$. According to the iteration termination conditions, we obtain $\hat Q_{min}\approx 0.5657$.
\begin{table}[htbp]
\centering
\caption{Iteration data of the repetitive electric Penrose process; the blue data in the last row no longer satisfy the iteration conditions.}
\label{tab:penrose}
\begin{tabular}{cccccccc}
\hline
$n$ & $\hat Q_n$ & $M_n/M_0$ & $E_{ extractable,n}/M_0$ & $E_{ extracted,n}/M_0$ & $M_{ irr,n}/M_0$ & $\xi_n$ & $\Xi_n$ \\
\hline
0 & 1.4142 & 1.0000 & 1.0000 & 0.0000 & 0.0000 & 0.0000 & 0.0000 \\
1 & 1.3945 & 0.9941 & 0.8285 & 0.0059 & 0.1655 & 0.5407 & 0.0347 \\
2 & 1.3743 & 0.9882 & 0.7551 & 0.0118 & 0.2331 & 0.5344 & 0.0480 \\
3 & 1.3537 & 0.9826 & 0.6982 & 0.0174 & 0.2844 & 0.5280 & 0.0577 \\
4 & 1.3327 & 0.9771 & 0.6500 & 0.0229 & 0.3270 & 0.5215 & 0.0656 \\
5 & 1.3112 & 0.9717 & 0.6076 & 0.0283 & 0.3641 & 0.5149 & 0.0722 \\
6 & 1.2893 & 0.9665 & 0.5692 & 0.0335 & 0.3972 & 0.5083 & 0.0779 \\
7 & 1.2669 & 0.9614 & 0.5341 & 0.0386 & 0.4272 & 0.5015 & 0.0829 \\
8 & 1.2441 & 0.9565 & 0.5016 & 0.0435 & 0.4548 & 0.4947 & 0.0874 \\
9 & 1.2209 & 0.9517 & 0.4713 & 0.0483 & 0.4804 & 0.4878 & 0.0913 \\
10 & 1.1972 & 0.9471 & 0.4429 & 0.0529 & 0.5042 & 0.4807 & 0.0949 \\
11 & 1.1731 & 0.9427 & 0.4161 & 0.0573 & 0.5265 & 0.4736 & 0.0982 \\
12 & 1.1485 & 0.9384 & 0.3908 & 0.0616 & 0.5476 & 0.4664 & 0.1011 \\
13 & 1.1235 & 0.9344 & 0.3669 & 0.0656 & 0.5675 & 0.4591 & 0.1037 \\
14 & 1.0981 & 0.9304 & 0.3441 & 0.0696 & 0.5863 & 0.4517 & 0.1061 \\
15 & 1.0723 & 0.9267 & 0.3225 & 0.0733 & 0.6042 & 0.4442 & 0.1082 \\
16 & 1.0460 & 0.9232 & 0.3019 & 0.0768 & 0.6213 & 0.4366 & 0.1101 \\
17 & 1.0194 & 0.9198 & 0.2822 & 0.0802 & 0.6375 & 0.4289 & 0.1117 \\
18 & 0.9923 & 0.9166 & 0.2635 & 0.0834 & 0.6531 & 0.4211 & 0.1132 \\
19 & 0.9649 & 0.9136 & 0.2457 & 0.0864 & 0.6679 & 0.4133 & 0.1145 \\
20 & 0.9371 & 0.9108 & 0.2286 & 0.0892 & 0.6822 & 0.4053 & 0.1156 \\
21 & 0.9089 & 0.9082 & 0.2124 & 0.0918 & 0.6958 & 0.3973 & 0.1165 \\
22 & 0.8803 & 0.9058 & 0.1969 & 0.0942 & 0.7089 & 0.3892 & 0.1173 \\
23 & 0.8515 & 0.9036 & 0.1821 & 0.0964 & 0.7215 & 0.3810 & 0.1179 \\
24 & 0.8223 & 0.9016 & 0.1681 & 0.0984 & 0.7335 & 0.3727 & 0.1183 \\
25 & 0.7928 & 0.8998 & 0.1547 & 0.1002 & 0.7451 & 0.3643 & 0.1185 \\
26 & 0.7629 & 0.8982 & 0.1419 & 0.1018 & 0.7563 & 0.3559 & 0.1186 \\
27 & 0.7329 & 0.8968 & 0.1298 & 0.1032 & 0.7670 & 0.3474 & 0.1186 \\
28 & 0.7025 & 0.8957 & 0.1183 & 0.1043 & 0.7773 & 0.3388 & 0.1183 \\
29 & 0.6719 & 0.8947 & 0.1074 & 0.1053 & 0.7873 & 0.3301 & 0.1180 \\
30 & 0.6411 & 0.8940 & 0.0971 & 0.1060 & 0.7968 & 0.3213 & 0.1174 \\
31 & 0.6101 & 0.8934 & 0.0874 & 0.1066 & 0.8060 & 0.3125 & 0.1168 \\
32 & 0.5790 & 0.8931 & 0.0783 & 0.1069 & 0.8149 & 0.3036 & 0.1160 \\
\textcolor{blue}{33} & \textcolor{blue}{0.5476} & \textcolor{blue}{0.8930} & \textcolor{blue}{0.0697} & \textcolor{blue}{0.1070} & \textcolor{blue}{0.8234} & \textcolor{blue}{0.2947} & \textcolor{blue}{0.1150} \\
\hline
\end{tabular}
\end{table}
From Table\ref{tab:penrose}, it can be seen that the iteration stops at the 32nd time. After the sequence ends, the return on investment of energy extraction is $30.36\%$, higher than the $20.7\%$ of the mechanical Penrose process in an extremal Kerr black hole\cite{11}. In addition, between two adjacent steps, only part of the decrease in extractable energy flows into the increase in extracted energy, while most flows into the increase in irreducible mass. Throughout the entire iteration sequence, the energy utilization efficiency is only $11.6\%$, i.e., only $11.6\%$ of the decrease in extractable energy is converted into extracted energy, while $88.4\%$ flows into the irreducible mass. This is due to the nonlinear behavior of the iterative process. In addition, after the iterations end, there remains $0.0783M_0$ of extractable energy to be extracted by other means. On the other hand, after the iterations end, the charge and mass of the black hole have not disappeared, which is similar to the repetitive Penrose process in a Kerr black hole\cite{8}.

\begin{figure}[!h]
  \centering
  \begin{subfigure}
    \centering
    \includegraphics[width=0.32\linewidth]{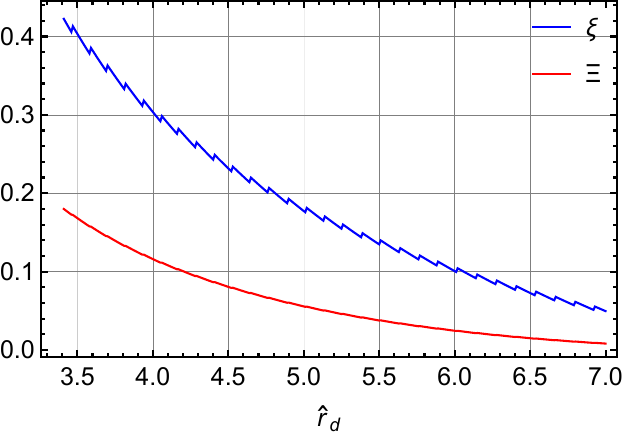}
  \end{subfigure}
  \begin{subfigure}
    \centering
    \includegraphics[width=0.32\linewidth]{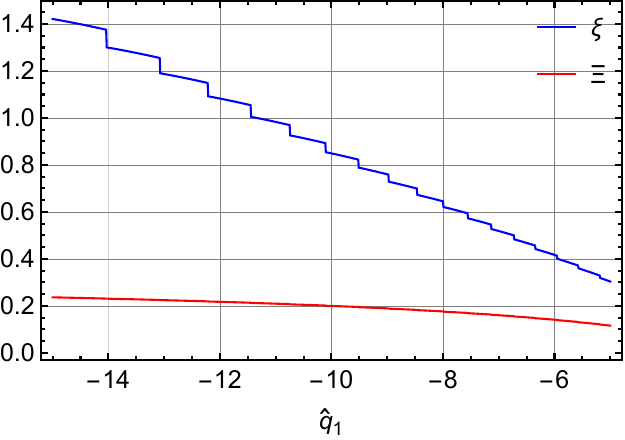}  
  \end{subfigure}
   \begin{subfigure}
    \centering
    \includegraphics[width=0.32\linewidth]{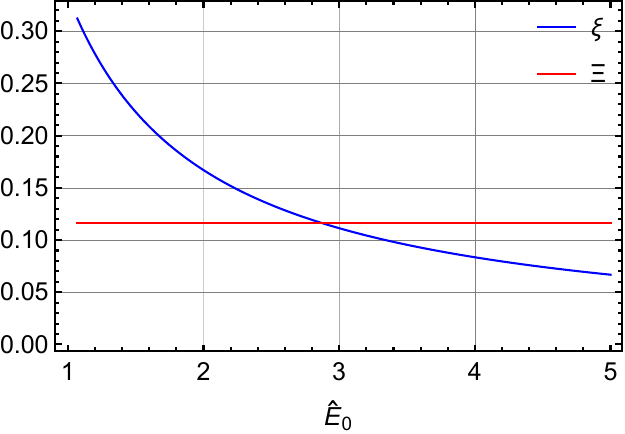}  
  \end{subfigure}
\caption{After the repetitive electric Penrose process ends, the return on investment of energy extraction and the energy utilization efficiency as functions of different decay radii $\hat r_d$, different charge-to-mass ratios $\hat q_1$ of particle 1, and different specific energies $\hat E_0$ of the incident particle. The oscillatory behavior of the curves arises because the iterative process is discrete.}
  \label{fig:2}
\end{figure}

In Fig.\ref{fig:2}, we show, after the iterations end, the return on investment of energy extraction and the energy utilization efficiency as functions of different decay radii $\hat r_d$, different charge-to-mass ratios $\hat q_1$ of particle 1, and different specific energies $\hat E_0$ of the incident particle. We choose the basic quantities $\hat r_d=4,\hat q_1=-5,\hat E_0=1.1$. When one of these variables needs to be changed, the other two are held fixed, and all other data are the same as in Table\ref{tab:penrose}.
From Fig.\ref{fig:2}, it can be seen that a smaller decay radius and a larger absolute value of the charge-to-mass ratio of particle 1 lead to a larger return on investment and energy utilization efficiency of energy extraction. The larger the specific energy of the incident particle, the smaller the return on investment of energy extraction, while the energy utilization efficiency remains unchanged. The return on investment of energy extraction can exceed $100\%$. However, the energy utilization efficiency can hardly exceed $50\%$, again indicating that the decrease in extractable energy is mainly converted into irreducible mass rather than extracted energy.

\begin{figure}[!h]
  \centering
    \includegraphics[width=0.45\linewidth]{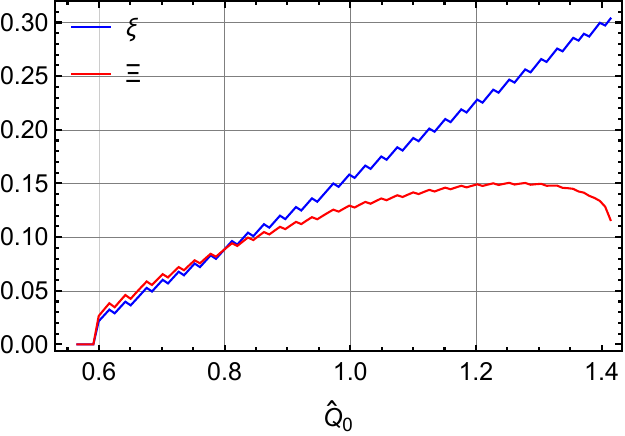} 
 \caption{For the case where the initial black hole is non-extremal, after the repetitive electric Penrose process ends, the return on investment of energy extraction and the energy utilization efficiency as functions of the initial black hole charge.}
\label{fig:3}
\end{figure}

In Table \ref{tab:penrose} and Fig.\ref{fig:2}, we have considered the case where the initial black hole is extremal. Therefore, in Fig.\ref{fig:3} we consider the case where the initial black hole is non-extremal. Except for the initial black hole charge, the parameters are exactly the same as in Table \ref{tab:penrose}.
It can be seen that as the initial black hole charge gradually increases and the black hole changes from non-extremal to extremal, the return on investment of energy extraction gradually increases, while the energy utilization efficiency first increases and then decreases.

\section{Conclusions}
This paper systematically studies the repetitive electric Penrose process in the GMGHS black hole. We first derive the radial motion equation and effective potential of particles in this black hole, and rigorously prove that the optimal condition for maximum energy extraction corresponds to the vanishing of the radial momenta of all three particles at the decay location, thereby giving the analytical solution of the basic equation\eqref{18}. Unlike the RN black hole case, the effective potential curve of the incident particle in the GMGHS black hole remains unchanged during the iterations, which originates from the special structure of its metric.

The iteration termination conditions are determined by two aspects: (i) the escape condition\eqref{23} requires that particle 0 can overcome the effective potential peak and escape to infinity, and this condition is independent of the black hole charge, which is still different from the RN black hole case; (ii) the existence condition\eqref{25} for the negative-energy particle 1 requires the black hole charge to be sufficiently large. The latter indicates that the repetitive process cannot completely consume the charge of the GMGHS black hole to zero, but by increasing the charge-to-mass ratio of incident particle 1, the remaining charge can be made arbitrarily small. This is qualitatively consistent with the conclusion for the RN black hole, but the quantitative form differs due to the different metric.

After the iterations end, the final return on investment of energy extraction can exceed $100\%$, indicating that it is feasible to obtain a large return with a small input. However, the energy utilization efficiency does not exceed $50\%$, indicating that most of the extractable energy is converted into an increase in the irreducible mass of the black hole, and only a small part is actually extracted. In addition, a smaller decay radius and a larger absolute value of the charge-to-mass ratio of particle 1 can improve the return on investment and energy utilization efficiency of energy extraction; the specific energy of the incident particle has a suppressive effect on the return on investment of energy extraction but does not affect the energy utilization efficiency; for an initially non-extremal black hole, as the initial charge increases, the return on investment of energy extraction increases monotonically, while the energy utilization efficiency first increases and then decreases. After the iterations end, the black hole still retains remaining extractable energy, and neither the charge nor the mass vanishes.

For future research, an important direction is to explore the co-evolution of charge and spin in the Kerr-Sen black hole\cite{12} under the repetitive electric Penrose process. Although Ref.\cite{15} studied the repetitive Penrose process in the Kerr-Sen black hole, that was the mechanical Penrose process.

\noindent {\bf Acknowledgments}

\noindent
This work is supported by the National Natural Science Foundation of China (Grants No. 12075143,12375043), and Natural Science Foundation of Shanxi
Province, China (Grant No.2025030221211241).

\end{document}